\documentclass[
               ]{jacow}

\usepackage[english]{babel}
\usepackage{hyperref}
\newcommand{\doi}[1]{\href{https://doi.org/#1}{https://doi.org/#1}}

\begin{document}

\title{Vector potential-based beam current sensors: applications for \NoCaseChange{Mu2e}, and picosecond bunch substructure}




\author{P. W. Gorham$^1$, M. T. Hedges$^2$, R. Kim$^2$, P. M. Lewis$^1$, K. Lynch$^2$, \\
        C. Miki$^1$, J. Morgan$^2$, N. Nyza$^1$, D. Pang$^1$, T. Polischuk$^1$, D. Still$^2$ \\
        $^1$Dept. of Physics \& Astronomy, Univ. of Hawaii at Manoa, Honolulu, HI, USA \\
        $^2$Fermi National Accelerator Laboratory, Batavia, IL, USA}

\maketitle

\begin{abstract}
We report on a new beam current monitor system based on work originally done to develop timing planes for future collider detectors. In contrast to sensors based on wall currents or toroid inductance, we make use of electric field induced by the electromagnetic vector potential of the beam current as it passes near or through a transmission line (TL), arranged with its dominant TEM or TE10 fields aligned with the current, thus directly coupling to TL modes. For resonantly extracted proton pulses currently being commissioned at the Fermilab Delivery Ring, we utilized a TEM parallel-plate TL geometry, since the requirements included measuring the structure of the bunch over $\sim{}200~\rm{ns}$ time scales. We present the design and results of simulated and realized sensors now deployed at Fermilab for future use as pulse-to-pulse proton intensity monitors at the Mu2e experiment. In addition, inspired by this and prior work, we investigate the use of broadband mm-wave dual-ridge waveguides as a means of measuring substructure of nanosecond bunches down to the tens of picosecond level for future use at the Long Baseline Neutrino Facility (LBNF).
\end{abstract}

\section{Introduction}
In this report we discuss two beam diagnostic applications, one of which has been implemented in an initial prototype with beam test results, for estimating the proton bunch current in the Mu2e experiment's 8~GeV bunches of $\sim 40 \times 10^6$ protons in $\sim 150$~ns~\cite{Bernstein:2019mu2e, Mu2e:2015tdr}, with several nanosecond resolution. The second application has relevance for the 120 GeV, $10^{11}$ protons, $\sim 1$~ns wide Fermilab Main Injector bunches for the Long Baseline Neutrino Facility (LBNF)~\cite{Tariq:2017lbnf}, and we develop and report on detailed electrodynamic simulations for reconstructing potential beam bunch microstructure at resolutions of 10-20~ps. 

The motivation for the approach in each case relies on an understanding of the evolution of the beam-current vector potential, which directly encodes the structure of the bunch passage, and which is the electrodynamic precursor of a radiated electric field in an appropriate sensor geometry. This approach naturally leads to sensor geometries that utilize simple transmission lines, and we implement these in both cases as realized detectors.

\section{Analytical basis}


We want to sense a wideband radiation field that depends directly on the beam current: 
$$\mathbf{E}(t) \propto \mathbf J_b(t)$$
and recall 
$$\mathbf{E} = -\nabla\Phi - {\partial{\mathbf {A}}}/{\partial t}$$ where $\Phi$ is the scalar potential and $\mathbf{A}$ is the vector potential. But the relativistic ``pancake'' E-field of the beam bunch is almost completely radial.
  
In the Coulomb gauge, 
$$\Phi_{\rm C}(\mathbf x,t) = \frac{1}{4\pi\epsilon} \int \frac{\rho(\mathbf x',t)}{|\mathbf x-\mathbf x'|}\,d^3x'$$
and  
$$\mathbf A_{\rm C} \sim \int \frac{ \mathbf J_b- \hat{\mathbf R}(\hat{\mathbf R}\cdot\mathbf J_b)}{R}\,d^3x'.$$
The vector potential in particular {\bf is} directly proportional to the integral beam current, and $\mathbf E(t)$ to $\partial{\mathbf A}/\partial{t}$.
  
But in the ideal case, beam current and wall current guarantee that the two potentials almost exactly cancel in the $z$ direction: 
$$(-\nabla\Phi - {\partial{\mathbf {A}}}/{\partial t})_z \simeq {1}/{\gamma^2}$$ 
and only the radial field is left as noted above.
  
We therefore need some way to perturb the wall current and break the beam pipe symmetry, without creating a large disturbance. Since we want to create an external radiation field, let's use an aperture in the wall. If that aperture is the opening to a transverse waveguide, then we can extract a field that propagates to an external sensor. A beam button monitor does something quite similar, although the fields are coupled via a capacitive pickup rather than radiative mode. 

\begin{figure}[!hb]
\centerline{~~\includegraphics[width=0.8\columnwidth, trim=0mm 0mm 0mm 0mm, clip]{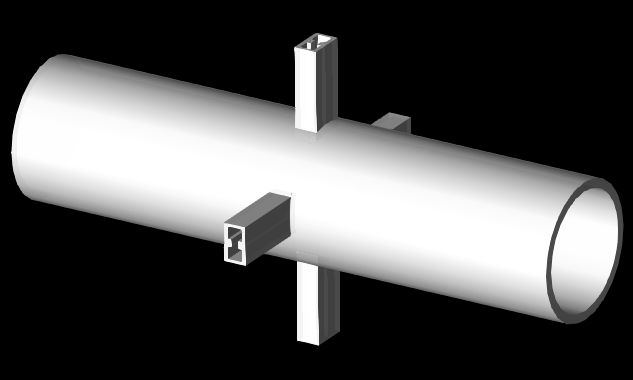}}
\caption{ Quad-ridged waveguide coupled to apertures in a beam pipe.
\label{beam2} }
\end{figure}
  
For rectangular waveguide (cf. Fig.~\ref{beam2}), we can align the dominant TE$_{10}$ mode with $A_z \propto J_z \simeq \mathbf J_b$. Then to first order the potential gradient is confined to the near field in the beampipe, and 
$$\mathbf E_{TE_{10}} = -\frac{\partial{\mathbf {A}}}{\partial t} \propto \mathbf{J_b}.$$ 
The waveguide modes then encode a signal proportional to the beam current via the vector potential. The one condition that must be imposed on this result is that the transfer function of the waveguide, including its low-frequency cutoff and its in-band delay dispersion, will be convolved with the beam current response, and thus the equation above will only apply within the waveguide passband, subject to deconvolution of transfer function effects.

Stepping back from these details, it should be noted that with a different choice of gauge, the role of the vector potential could change or even largely disappear, and neither the vector or scalar potential are themselves directly measurable, except through the fields they produce. Thus this approach is intended primarily to foster a better intuition about the relation between the vector potential and the current --- because it is proportional to an integral over the current density, and the radiated electromagnetic field depends on the derivative of that potential, it does uniquely highlight the direct coupling between beam current and the radiated field used to sense it.

\section{\NoCaseChange{Mu2e} Application}

Applying the methods outlined above to measurement of the Mu2e proton beam current is complicated by one significant change. In our application, the goal is to measure the portion of the beam current that does not interact with the production target, and which travels beyond the target into the dump through a several meter air gap. Thus the bunches will interact with the containment vessel wall, and with a small burden of air, and will also be subject to the extreme radiation in proximity to the dump. During a spill, bunches of $40 \times 10^6$ 8~GeV protons arrive every 1.695 microseconds for several tens of milliseconds, and these spills occur 8~times per 1.4 s cycle continuously over a period of several years.

\begin{figure}
\vspace{-0mm}
\centerline{~~\includegraphics[width=0.8\columnwidth, trim=0mm 0mm 0mm 0mm, clip]{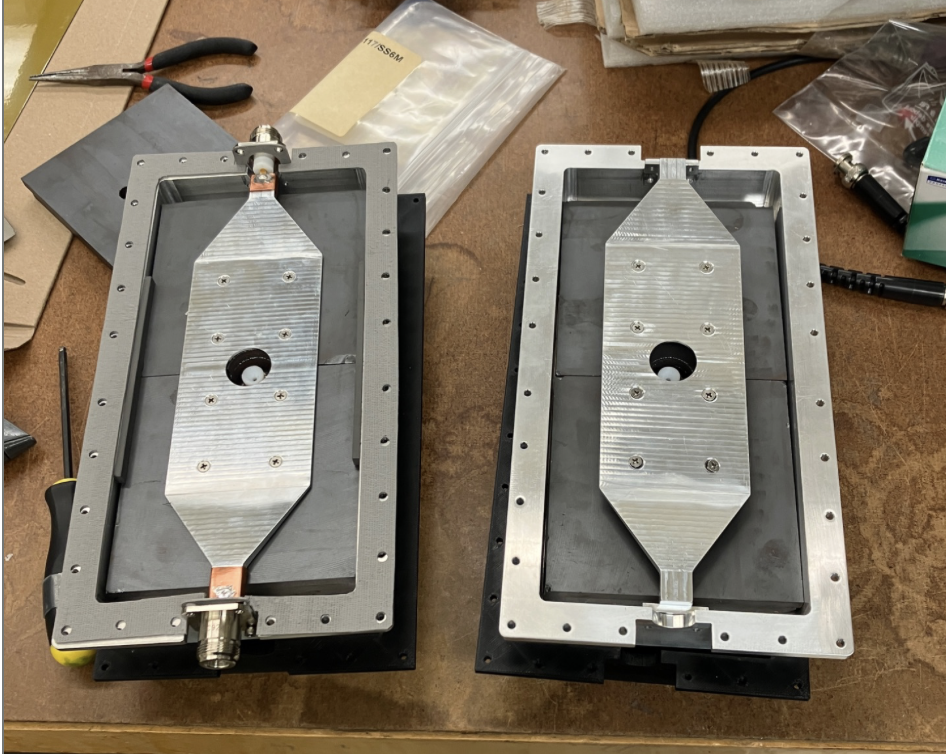}}
\caption{ Mu2e sensor in its split clamshell housing with ferrite for cavity mode suppression. The 50 $\Omega$ TEM parallel plate waveguide is formed when the two halves are joined.
\label{mACE1} }
\end{figure}

The detector must therefore observe the beam current in a system which does not include the beam pipe or the wall current. The expected peak current during a $\sim 150$~ns bunch passage is in the microampere range, and thus is a challenge for any traditional beam current monitor, at least in standard configurations. As such it will require low-noise amplification to raise the induced signal to levels that can be read out by standard analog-to-digital converters (ADC). And since the environment itself will include RF backgrounds far above the induced radiation fields, the sensor must be enclosed in a Faraday cage that completely isolates it from such backgrounds.

Figure~\ref{mACE1} shows the prototype sensor, denoted as mACE (for mu2e Askaryan Calorimeter Element), we have developed for this application. It consists of two aluminum clamshells that join to form a rectangular box, with a 50~$\Omega$ parallel-plate TEM waveguide running along its long axis. The aperture through the waveguide center allows for a calibration current element (a wire) to be inserted. The space above and below the waveguide is loaded with ferrite, Fairrite type 42 $100\times{}100\times{}6.4~\rm{mm}$ blocks, in four 3-element stacks to suppress cavity resonance modes within the Faraday enclosure. N-type RF bulkhead connectors extract the RF signal.

\begin{figure}
\vspace{-0mm}
\centerline{~~\includegraphics[width=0.8\columnwidth, trim=0mm 0mm 0mm 0mm, clip]{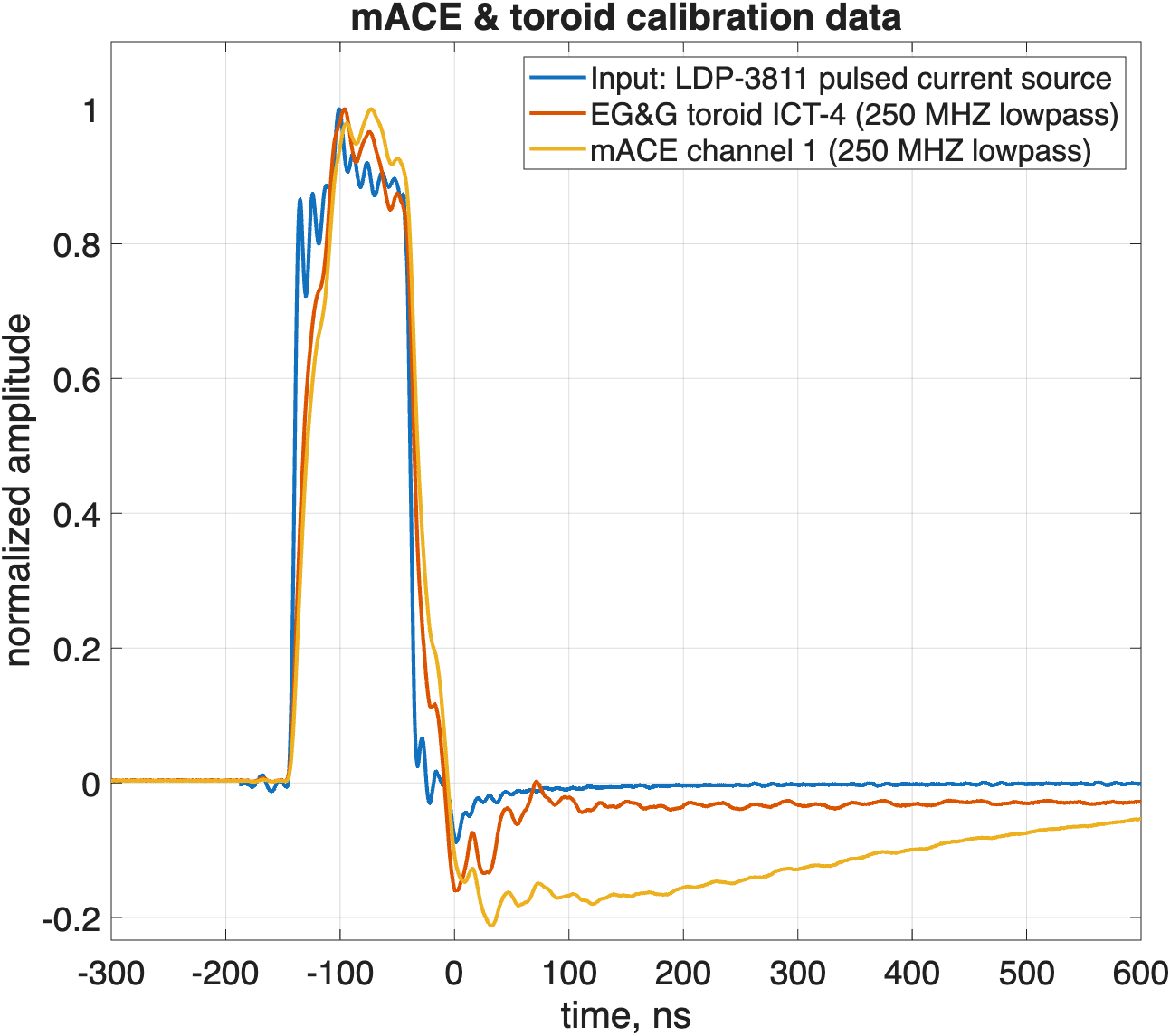}}
\caption{Calibration data for Mu2e sensor compared to input current pulse and the response of a high-fidelity, high-bandwidth toroid transformer developed by EG\&G. 
\label{mACEcal} }
\end{figure}

To validate performance, we compared the mACE sensor to a custom high-fidelity, high-bandwidth integrating current transformer (ICT) based on a ferrite-loaded toroid developed for accelerator applications by EG\&G. The results of the calibration data are shown in Fig.~\ref{mACEcal}. Both the mACE sensor and the ICT show some degree of ``droop'' (overshoot followed by a return to baseline) compared to the input signal, though it is larger by factors of 2-3 for the mACE sensor. Such effects are common in the transfer function for low-frequencies, and are correctable in post-analysis.

\subsection{Beam-Test Results}
We performed initial beam tests of the mACE prototype during late May and early June 2026, with the sensor positioned  with a several meter air gap past a window in the end of the M4 muon-campus line. The EG\&G ICT-4 was positioned upstream of it.  Well upstream of both, near the entrance of the M4 line, a wall-current monitor (WCM) made measurements of the beam current signal as it entered the line, with an 8 ns sample rate. The mACE and ICT-4 were sampled at both a 2 ns rate (500 Ms/s) and separately with a 0.5 ns rate (2 Gs/s). 

\begin{figure}
\vspace{-0mm}
\centerline{~~\includegraphics[width=\columnwidth, trim=0mm 0mm 0mm 0mm, clip]{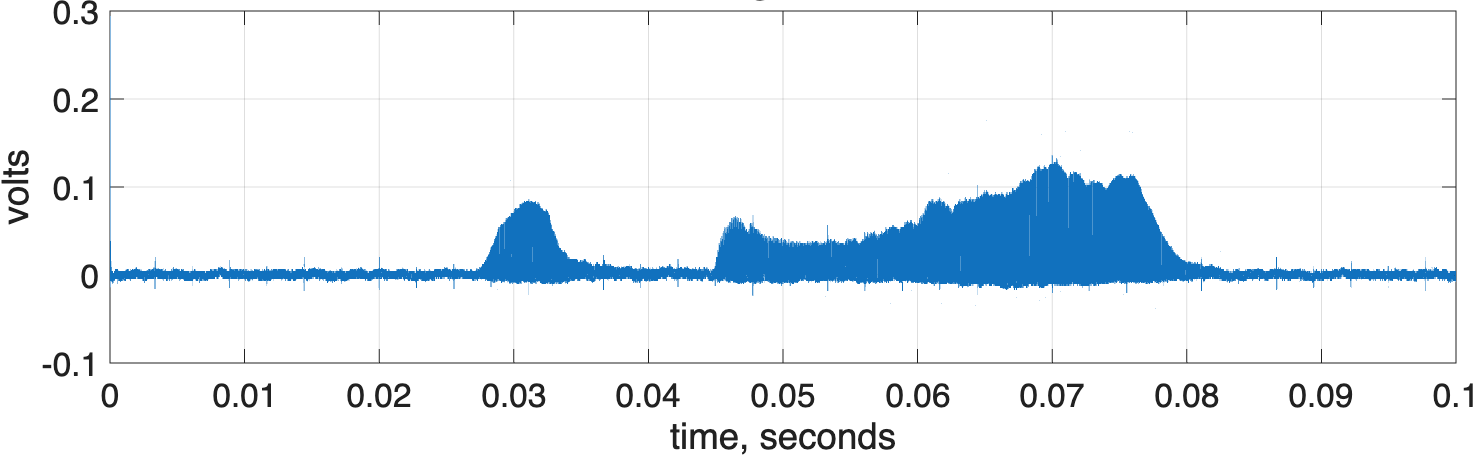}}
\centerline{~~\includegraphics[width=\columnwidth, trim=0mm 0mm 0mm 0mm, clip]{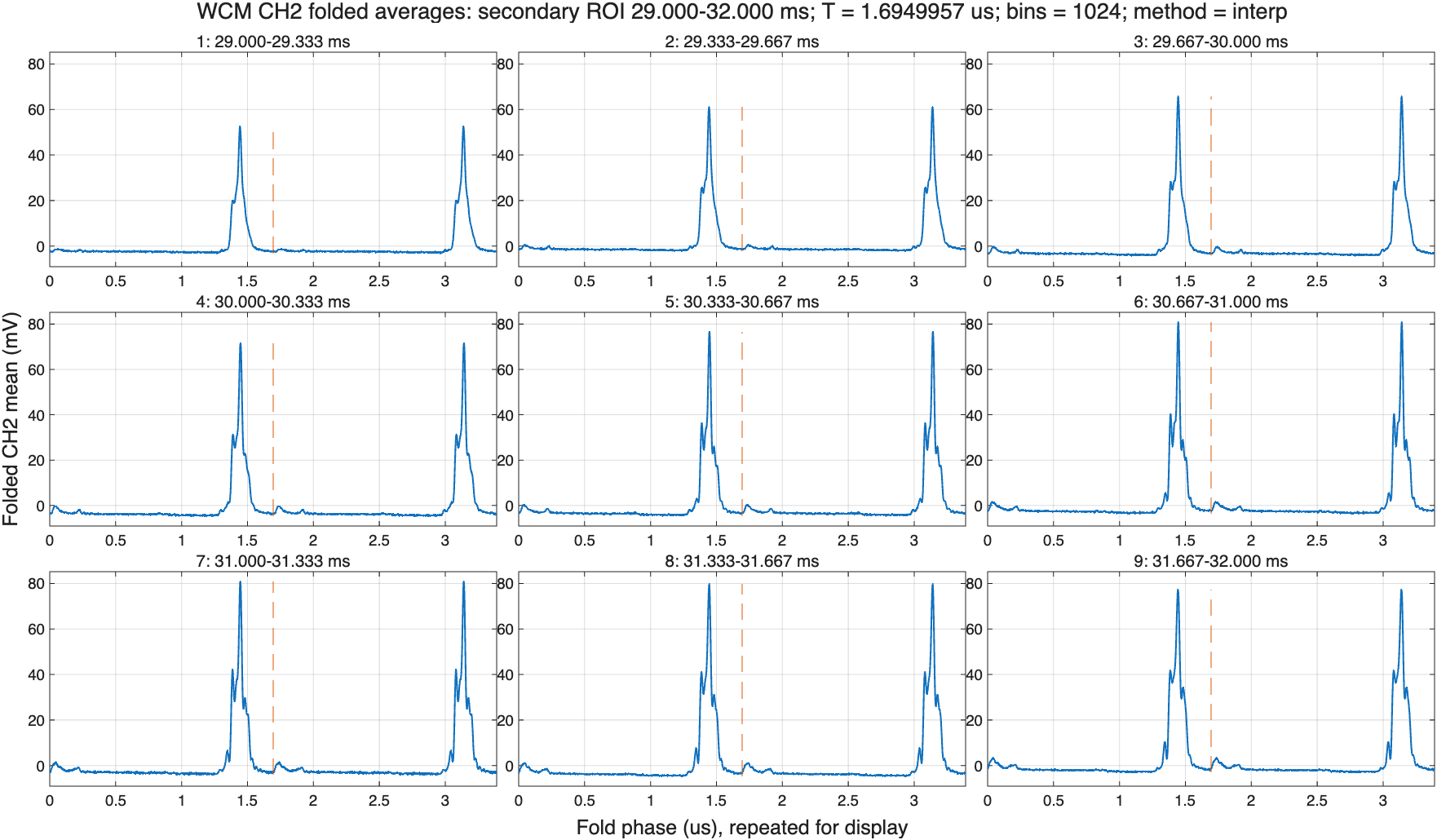}}
\centerline{~~\includegraphics[width=\columnwidth, trim=0mm 0mm 0mm 0mm, clip]{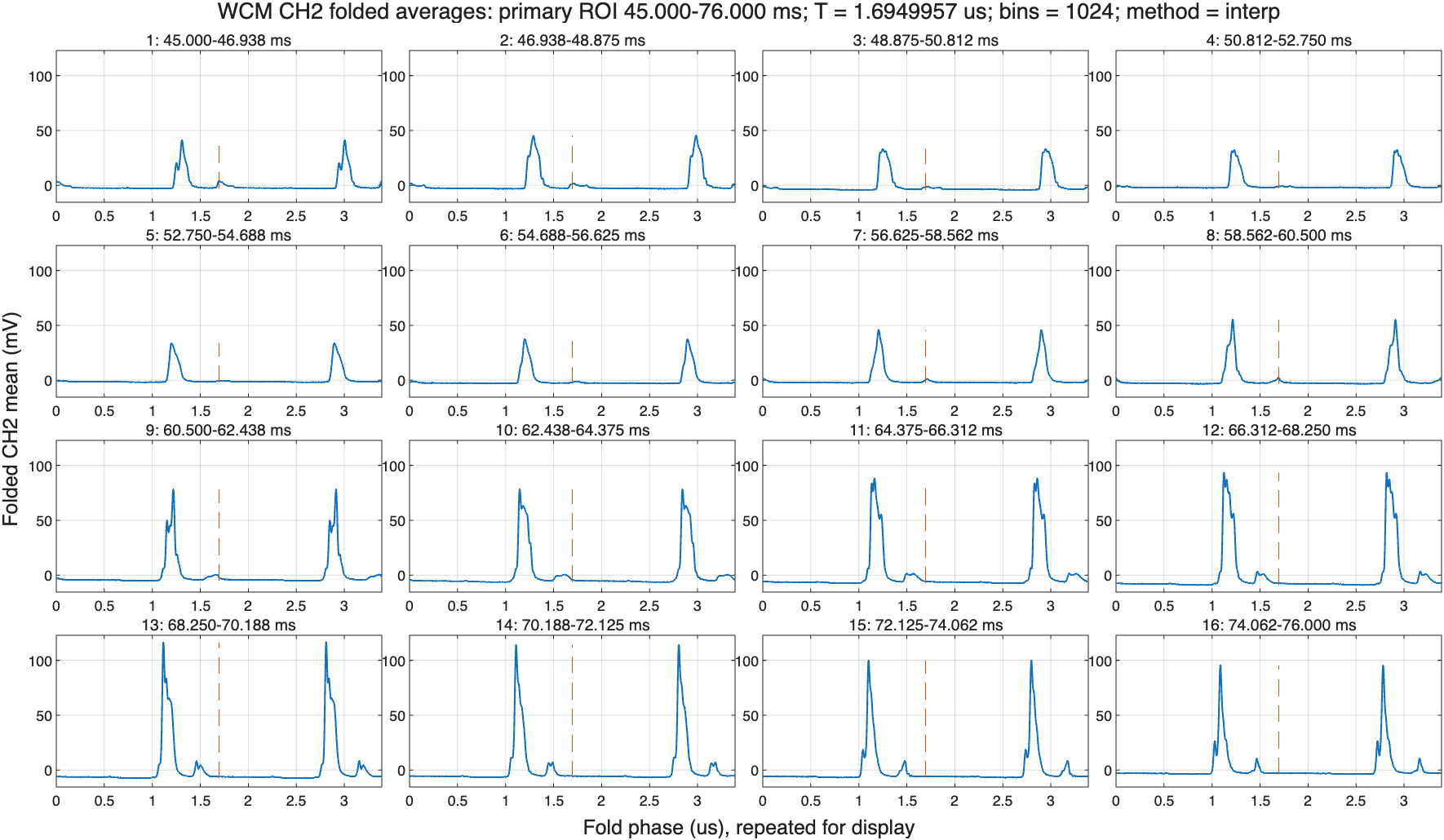}}
\caption{ Top: Wall current monitor data for the entire spill: the sextupole instability, followed by the main spill, on a wide scale; Middle: zoom of several average pulse shapes in the sextupole instability; Bottom: Zoom of several average pulse shapes for the main spill.
\label{WCMdata} }
\end{figure}

Figure~\ref{WCMdata} shows several views of the WCM data, both an overall view of the entire spill (top), which is a sequence of bunches recurring at 1.695~$\upmu$s periodicity, including a ``pre-spill" due to sextupole instability, (at around 30 ms) and the main spill from 45-80 ms, and a view of average pulse shapes within several locations during the sextupole instability (middle) and main spill (bottom). The bunch durations are typically $\lesssim 150$~ns, with variable sharp features appearing in various phases of the bunch pulse throughout the spill.

\begin{figure}
\vspace{-0mm}
\centerline{\includegraphics[width=\columnwidth, trim=0mm 0mm 0mm 0mm, clip]{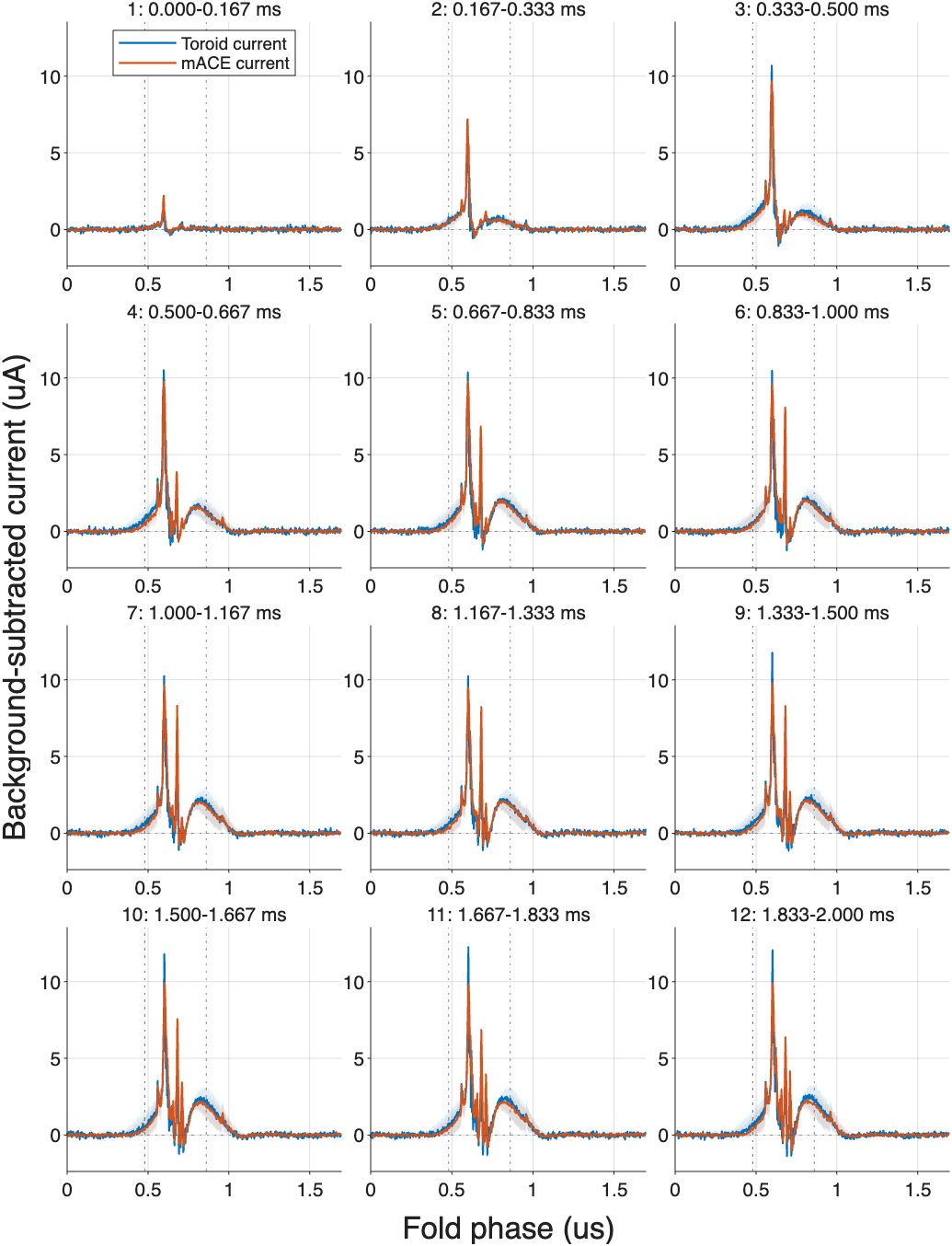}}
\caption{ Average pulse profiles measured with the mACE sensor and the ICT-4 commercial toroid, overlain, for various segments of the complete Mu2e spill.
\label{mACE_toroid1} }
\end{figure}

Examples of preliminary mACE and ICT-4 toroid measurements of pulse structure within the main  portion of the M4 line spill are shown in Fig.~\ref{mACE_toroid1}. These are again averaged profiles within a given portion of the main spill, and we have subtracted a slowly varying background level from each plot, and converted these to microamperes using our calibration results. The two outputs were not forced to the same level; in fact the calibration was precise enough that the resulting curves are highly commensurate in the absolute current levels. However, the pulse structure is subject to beam backgrounds that are not yet well understood, and thus the broader structure of the pulse is still under investigation. Fine structure down to the several ns level has been measured here and throughout other portions of the run, confirming that the sensors retained their full effective bandwidth. Whereas the WCM saw peak currents of around 6 microamperes, mACE and our fast toroid observe peaks nearly twice that level, indicating that the high-resolution sampling can reveal important details.

\section{LBNF/Main Injector application}

At an order of $10^{11}$ or more protons per nanosecond-scale bunch, proton shot noise is negligible, and the secondary production in the target for the LBNF to DUNE beamline will not depend on beam structure to first order.
The most likely deviations from the idealized smooth gaussian would be coherent effects caused by RF manipulation, phase noise, nonadiabatic transitions, slip-stacking remnants, longitudinal filamentation, or impedance-driven instability. In simulations of proposed 531 MHz rebunching, Fermilab researchers specifically noted that kinks and nonadiabatic changes in RF voltage can create unwanted longitudinal harmonics and broaden the resulting bunches. These effects may also induce microstructure in the bunch that are currently unmeasured at the finest spatial scales. A fast pickup close to extraction or the target could therefore detect microstructure associated with:
parasitic longitudinal harmonics,
satellite sub-bunches,
extraction-phase jitter,
bunch-to-bunch charge variation,
tails outside the intended RF bucket, and
pulse-to-pulse changes associated with upstream instability.

\begin{figure}
\vspace{-0mm}
\centerline{\includegraphics[width=\columnwidth, trim=0mm 0mm 0mm 0mm, clip]{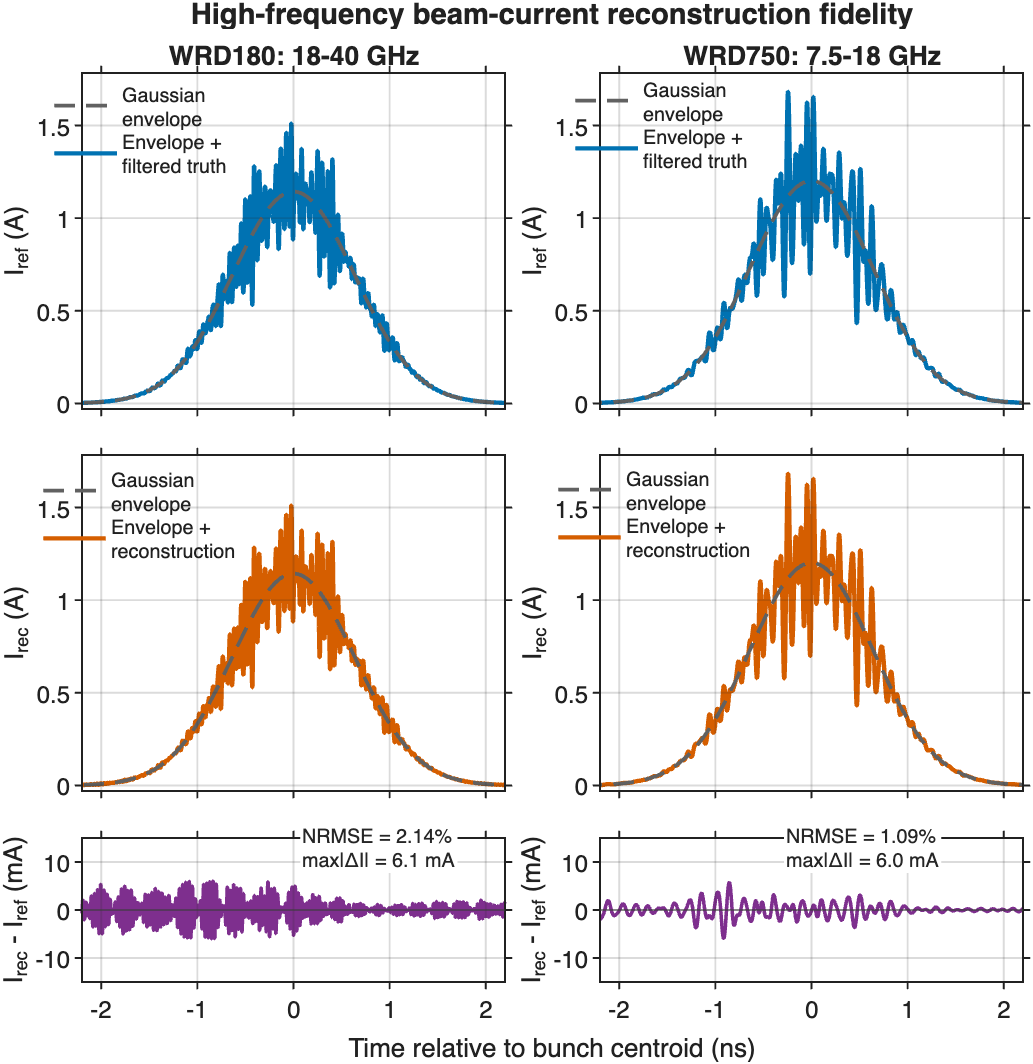}}
\caption{ Simulated reconstruction of beam current microstructure over two complementary double-ridged waveguide passbands, WRD180 (left column) and WRD750 (right). The top row shows the input gaussian beam current with superposed microstructure (filtered to the same passband), the middle row are the reconstructions, and the bottom row gives the residuals.
\label{bcrecon} }
\end{figure}

There is a particularly interesting LBNF application that makes the proton time profile a genuine physics observable, not merely an accelerator diagnostic. The proton RF time structure is imprinted on the secondary hadrons and therefore on the neutrino beam. Lower-energy parent pions and kaons generally produce later-arriving neutrinos than more energetic parents. Time-slicing neutrino events relative to the parent proton bunch therefore selects different neutrino-energy and flavor distributions. This is the proposed ``stroboscopic'' approach~\cite{Angelico:2019pyo}. The difficulty is that the ordinary $\sim 1$ ns Main Injector bunch width largely washes out the time–energy correlation. Studies find that proton bunches approaching 100 ps, together with comparable detector timing, would make the method substantially more useful. A spill-by-spill, causal reconstruction of $I_p(t)$ could then enter directly into the event likelihood or flux prediction. Width, tails, satellite bunches, asymmetry, and timing jitter would all be relevant. 

Figure~\ref{bcrecon} shows results of a full-wave Finite Difference Time Domain simulation using the quad-ridge waveguide beampipe ports (a single port of the four in this case) as shown in Fig.~\ref{bcrecon}, for a 1~ns LBNF-like bunch with superposed microstructure, filtered to match the passband of the waveguide. We took the raw waveguide output and deconvolved it with a measured impulse response for the system. In each case the microstructure is recovered to the 1-2\,\% level, under the assumption that a separate, low-frequency device such as a WCM or similar, provides the baseline gaussian truth measurement.

\section{Conclusions}

We have developed two distinct but related beam-current diagnostic methods, both based on the transfer of beam current information to a transmission line, which can be viewed as an application of vector potential theory under the Coulomb gauge. When a beam current passes through or near a transmission line, fields consistent with the dominant TEM or TE modes are induced with first-order coupling to the beam current, and we have demonstrated this for a Mu2e application, including results of an initial beam test with a prototype detector. We have also developed a high-fidelity full-wave electrodynamic simulation of the sensing of beam microstructure via a waveguide coupled to a beampipe aperture, and have demonstrated accurate reconstruction of the microstructure through deconvolution of the resulting waveguide output. While such methods certainly draw on a rich heritage of related methodologies such as WCMs and button monitors, they also provide a novel approach with its own potential richness for beam diagnostics development.


\end{document}